# Imaging the Fano lattice to 'hidden order' transition in URu$_2$Si$_2$

A. R. Schmidt[1,2], M. H. Hamidian[1,2], P. Wahl[1,3], F. Meier[1], A. V. Balatsky[4], J. D. Garrett[5], T. J. Williams[6], G. M. Luke[6,7] & J. C. Davis[1,2,8,9]

[1]Laboratory for Atomic and Solid State Physics, Department of Physics, Cornell University, Ithaca, New York 14853, USA. [2]Condensed Matter Physics and Materials Science Department, Brookhaven National Laboratory, Upton, New York 11973, USA. [3]Max-Planck-Institut für Festkörperforschung, Heisenbergstraße1, D-70569 Stuttgart, Germany. [4]Theory Division and Center for Integrated Nanotechnology, Los Alamos National Laboratory, Los Alamos, New Mexico 87545, USA. [5]Brockhouse Institute for Materials Research, McMaster University, Hamilton, Ontario, L85 4M1, Canada. [6]Department of Physics and Astronomy, McMaster University, Hamilton, Ontario, L8S 4M1, Canada. [7]Canadian Institute for Advanced Research, Toronto, Ontario, M5G 1Z8, Canada. [8]School of Physics and Astronomy, University of St Andrews, St Andrews, Fife KY16 9SS, UK. [9]Department of Physics and Astronomy, University of British Columbia, Vancouver, V6T 1Z1, Canada.

**Within a Kondo lattice, the strong hybridization between electrons localized in real space (*r*-space) and those delocalized in momentum-space (*k*-space) generates exotic electronic states called 'heavy fermions'. In URu$_2$Si$_2$ these effects begin at temperatures around 55 K but they are suddenly altered by an unidentified electronic phase transition at $T_o$ = 17.5 K. Whether this is conventional ordering of the *k*-space states, or a change in the hybridization of the *r*-space states at each U atom, is unknown. Here we use spectroscopic imaging scanning tunnelling microscopy (SI-STM) to image the evolution of URu$_2$Si$_2$ electronic structure simultaneously in *r*-space and *k*-space. Above $T_o$, the 'Fano lattice' electronic structure predicted for Kondo screening of a magnetic lattice is revealed. Below $T_o$, a partial energy gap without any associated density-wave signatures emerges from this Fano lattice. Heavy-quasiparticle interference imaging within this gap reveals its cause as the rapid splitting below $T_o$ of a light *k*-space band into two new heavy fermion bands. Thus, the URu$_2$Si$_2$ 'hidden order' state emerges directly from the Fano lattice electronic structure and exhibits characteristics, not of a conventional density wave, but of sudden alterations in both the hybridization at each U atom and the associated heavy fermion states.**

When a spin-half electronic state is embedded in a metal, the Kondo effect[1–4] screens it, generating a spin-zero many-body state (Fig. 1a). This state can be imaged at magnetic adatoms on metal surfaces[5–7] by using SI-STM techniques in which the tip-sample differential tunnelling conductance $dI/dV(\mathbf{r}, E) \equiv g(\mathbf{r}, E)$ is proportional to the local density of electronic states LDOS($\mathbf{r}$, $E$). However, there are two possible channels for these tunnelling electrons (either to $\mathbf{r}$-space many-body states or to $\mathbf{k}$-space delocalized states), so the measured conductance $g(\mathbf{r}, E)$ is best parameterized as a two-channel Fano spectrum[5] (Fig. 1b):

$$g(\mathbf{r}, E) \propto \frac{(\varsigma + E')^2}{E'^2 + 1} \quad \text{where} \quad E' = \frac{(E - \varepsilon_0)}{\Gamma/2} \tag{1}$$

Here $E$ is the electron energy, $\varsigma$ is the ratio of the probabilities of tunnelling into the $\mathbf{r}$-space many-body state or into the delocalized $\mathbf{k}$-space states, $\varepsilon_0$ is the energy of the many-body state and $\Gamma$ is proportional to its hybridization strength.

A periodic array of such atoms in a crystal lattice, each having $\mathbf{r}$-space states, the spins of which are screened by hybridization with the $\mathbf{k}$-space bands, is often referred to as a Kondo lattice[1–3]. Eventually as $T$ approaches zero, this hybridization process generates the famous 'heavy fermion' electronic states[1,8–11]. In the simplest picture, an initial light $\mathbf{k}$-space band $E_k$ is split by such hybridization into two heavy bands:

$$E_k^{\pm} = \frac{\tilde{\varepsilon}_k^f + E_k \pm \sqrt{\left(\tilde{\varepsilon}_k^f - E_k\right)^2 + 4|\tilde{V}_k|^2}}{2} \tag{2}$$

Here $\tilde{\varepsilon}_k^f$ is the renormalized energy of the $f$-states after they establish coherence, while the strength of their hybridization processes is represented by $\tilde{V}_k$ such that $\Gamma = \pi N(E_F) \left\langle \left|\tilde{V}_k\right|^2 \right\rangle$, where $N(E_F)$ is the density of states at the Fermi level (see chapter 10 of ref. 1). The type of band-structure in equation (2), which is a signature of such $\mathbf{r}$-space/$\mathbf{k}$-space hybridization, is shown schematically in Fig. 1c for a hole-like $E_k$. We note how the two heavy bands become widely separated in $\mathbf{k}$-space from the original light band (dashed line) where the avoided crossing occurs within the range of hybridization (arrows in Fig. 1c). No phase transition is expected with falling $T$ because this band structure emerges smoothly from the Kondo screening of the magnetic lattice.

For real materials, the Kondo lattice screening processes must be more intricate because the initial unscreened $\mathbf{r}$-space state can exhibit a complex spin manifold and its screening can be due to multiple bands. But despite these complexities, a heuristic hypothesis for SI-STM observables in a Kondo lattice might include a generalization of the Fano spectrum of equation (1) (Fig. 1b) to a 'Fano lattice'—an array of such Fano spectra with the same periodicity as the atoms the localized states of which are undergoing the hybridization. Indeed, where the **r**-space electronic characteristics of atoms undergoing such Kondo screening has been predicted using either dynamical mean-field theory[12] or large-$N$ expansion theory[13,14], the electronic structure can exhibit characteristics of such a Fano lattice.

## The hidden-order state of URu$_2$Si$_2$

After the Kondo screening of the magnetic sublattice has begun to alter the electronic structure of URu$_2$Si$_2$, the unexpected phase transition is signified by sudden changes at $T_o$ = 17.5 K in bulk properties such as resistivity, magnetic susceptibility and specific heat[15–17], in the **k**-space electronic structure[18], and by the emergence of a partial gap in the optical conductivity[19,20] and point contact measurements[21,22]. The new phase therefore is not 'hidden'—rather, it is the identity of its order parameter that is unknown. At $T \ll T_o$, this phase is a heavy fermion system with an effective electron mass $m^* \approx 25 m_e$ (refs 15, 16) that eventually becomes superconducting below 1.5 K. A transition to an antiferromagnetic state also occurs at $T_N \approx 17.5$ K, but neutron-scattering experiments reveal[23] that the ordered moment is far too small ($0.03 \mu_B$, where $\mu_B$ is the Bohr magneton) to account for the large entropy loss at $T_o$. Advanced experimental techniques have recently improved our understanding of the changes in the electronic and magnetic excitations at $T_o$. The spin excitations[24] above $T_o$ are gapless and overdamped at incommensurate wavevectors. Below $T_o$, sharp magnetic modes[25] appear near $(1 \pm 0.4, 0)\pi/a_0$ above a $\omega \approx 4$ meV energy gap. And, although angle-resolved photoemission cannot determine empty-state or **r**-space electronic structure, it does reveal[26] that, for the filled states above $T_o$, a light band crosses $E_F$ near $(0, \pm 0.3)\pi/a_0$; $(\pm 0.3, 0)\pi/a_0$ while, below $T_o$, this evolves into a heavy band below $E_F$.

Thus, the screening by light Ru-based $d$-electron bands of the $f$-electrons at each U atom apparently begins to alter the URu$_2$Si$_2$ electronic structure somewhere near $T \approx 55$ K (refs 15, 16). Then, however, the transition at $T_o$ = 17.5 K dramatically alters the density of states at the Fermi level[15,16,19,23], the electronic band structure[26] and the spin excitation spectrum[23,24]. Among the theoretical proposals to explain this transition to a hidden-order state are two basic classes. The first class considers **k**-space susceptibilities of the Fermi surface to a conventional density wave state[15,25,27–29], while the second class considers ordering of the **r**-space states at U atoms, with the corresponding alteration (via hybridization changes) to the band structure[12,30–34]. However, despite two decades of research, which of these classes correctly describes the hidden-order transition in URu$_2$Si$_2$ remains unknown.

**Conventional density wave versus *d-f* hybridization change**

In conventional metallic systems, the opening of an energy gap $\Delta$ below some critical temperature usually occurs as a result of a diverging susceptibility of the **k**-space states surrounding $E_F$. The pre-existing band structure is unaltered except within a few $\pm\Delta$ of $E_F$. Exemplary of such transitions are those to Peierls charge-density-wave or spin-density-wave states[35]. Here **r**-space modulations in charge and/or spin appear at a fixed wavevector **Q*** — which is the same for empty and filled gap-edge states, and the associated gap always spans $E_F$. In Kondo lattice systems, in contrast, the opening of a gap near $E_F$ is due to a profoundly different phenomenon[1–3,11]. Hybridization between the **r**-space and **k**-space electrons splits a light band into two heavy bands approximately separated in energy by the hybridization range (~ $\tilde{V}_k$ in equation (2)) and segregated widely in **k**-space (Fig. 1c). To understand the hidden-order transition in URu$_2$Si$_2$ we therefore face an elementary issue: does it occur because of a diverging Fermi-surface susceptibility of the **k**-space heavy fermions towards a conventional ordered state, or because of a sudden alteration of the many-body **r**-space states at each U atom with the associated changes in hybridization? To distinguish between these two fundamentally different situations requires determination of the electronic structure as temperature is reduced to $T_0$.

**SI-STM studies of URu$_2$Si$_2$**

For this purpose, we introduce SI-STM techniques to examine simultaneously the **r**-space/**k**-space electronic structure evolutions in URu$_2$Si$_2$. Here the tip-sample differential tunnelling conductance $dI/dV(\mathbf{r}, V) \equiv g(\mathbf{r}, E = eV)$ at locations **r** and sample-bias voltage $V$ yields an image proportional to the local density of electronic states LDOS(**r**, $E$), where $e$ is the electron charge. Moreover, **k**-space electronic structure elements can be determined simultaneously by using Fourier transform scanning tunnelling spectroscopy[36–42]. The pure compound URu$_2$Si$_2$ exhibits one excellent cleave surface that is Si-terminated, whereas the alloy U$_{0.99}$Th$_{0.01}$Ru$_2$Si$_2$, which we use for quasiparticle interference studies, exhibits a different excellent cleave surface that is U-terminated. Each of these two types of cleave surface has the same lattice periodicity and orientation, which is highly distinct from that of the Ru layer (inset in Fig. 2b). They can be distinguished from each other because the U-termination layer exhibits the correct density of impurity states generated by the 1% of U sites upon which Th atoms are substituted (Fig. 3a). In Fig. 2a we show a typical Si-terminated topographic image. Figure 2b shows, as open squares, the typical spatially averaged $\langle g(E) \rangle$ spectrum on such surfaces at 19 K where $T > T_o$. The red line is a fit of $\langle g(E) \rangle$ by equation (1), revealing its excellent parameterization as a Fano spectrum[5] (Supplementary Fig. 1). In Fig. 2c–e we show the atomically resolved images of the parameters of the Fano spectrum. Here $\varepsilon_0(\mathbf{r})$, $\Gamma(\mathbf{r})$, and $\zeta(\mathbf{r})$ are determined from fitting $g(\mathbf{r}, E = eV)$ for each pixel within the yellow box in Fig. 2a. Significantly, we find that the maximum in both $\varepsilon_0(\mathbf{r})$ and $\zeta(\mathbf{r})$ and the minimum in $\Gamma(\mathbf{r})$ occur at the U sites (X in Fig. 2c–e), as would be expected for a Kondo lattice of U atoms. These observations, in combination with theoretical predictions for such a phenomenology[12–14], indicate that the **r**-space 'Fano lattice' electronic structure of Kondo screening in magnetic lattices can now be visualized.

**Evolution of density of states at Si- and U-termination surfaces**

For U-terminated surfaces (see Fig. 3a), the spatially averaged density of states DOS($E$) ∝ <$g(E)$> spectrum for $T > T_o$ is less structured than that of the Si-terminated surface in Figs 2a and 3c. Typical <$g(E)$> spectra are shown as open squares for each listed temperature between 18.6 K and 1.9 K in the inset to Fig. 3a, with the top spectrum being characteristic of $T > T_o$. Upon cooling through $T_o$, strong changes are detected in the DOS($E$) in a narrow energy range (inset to Fig. 3a). By subtracting the spectrum for $T > T_o$, we determine how the DOS($E$) modifications due to the hidden-order state emerge rapidly below $T_o$ (Fig. 3b). They are not particle–hole symmetric, with the predominant effects occurring between –4 meV and +3 meV. For the Si-terminated surfaces upon cooling below $T_o$, the overall Fano lineshape of DOS($E$) as discussed in Fig. 2c–e is unchanged (Fig. 3c). In the inset to Fig. 3c, we show the evolution of the <$g(E)$> spectrum between 19 K and 1.7 K. In each case, the red line is the fit to the Fano spectrum at each temperature (excluding the data points in the bias range –7.75 mV to 6.75 mV) while the measured <$g(E)$> spectra are shown as open squares. Again, by subtracting the fitted Fano spectrum from the <$g(E)$> at each $T$ value we determine the temperature dependence of the hidden-order DOS($E$) modifications (Fig. 3d). At no $E$ value on either surface do these DOS($E$) spectra represent a complete gap. Finally, no changes are observed in the high-energy DOS($E$) as the temperature falls below $T_o$ (Supplementary Fig. 2), perhaps indicating that the basic Brillouin zone geometry is not altered by the transition.

**Heavy *f*-electron quasiparticle interference imaging**

To determine the evolution of **k**-space electronic structure through $T_o$, we use heavy-electron quasiparticle interference[43–45] (QPI) imaging. The Si-terminated surface has proved unproductive for this purpose because the Fourier transform of its $g(\mathbf{r}, E)$ images (Supplementary Fig. 3) are so complex that the multiple bands cannot yet be disentangled. However, in recent studies of heavy-fermion QPI in $Sr_3Ru_2O_7$ it was shown that replacing 1% of the Ru atoms by Ti atoms produced intense scattering interference and allowed successful **k**-space determination[43]. Emulating this approach, we substituted 1% Th atoms on the U sites, which results in crystals usually cleaving at the U layer. The average spectrum on this U-terminated surface develops the narrow resonant $DOS(E)$ structure below the $T_o$ (red data between vertical arrows in Fig. 3b), within which we observe intense QPI; see the $g(\mathbf{q}, E)$ movies in the Supplementary Information. The 1% Th substitution suppresses $T_o$ by only ~1 K (refs 46, 47) and does not alter the basic hidden order phenomenology (refs 46, 47), so the phenomena we report are not caused by our dilute Th doping. Moreover, because the energy scale of $DOS(E)$ alterations is consistent with Th-doped specific heat measurements[46] and because these alterations are already detectable in tunnelling within 1 K below the bulk transition (blue line in Fig. 3b), the electronic structure of the U-terminated surface appears to be bulk representative of the hidden-order phase.

For QPI studies of the hidden-order transition we therefore measure $g(\mathbf{r}, E = eV)$ in a 50 nm × 50 nm field of view (FOV) with 250 µV energy resolution and atomic spatial resolution on these U-terminated surfaces (the simultaneous topograph is shown in Supplementary Fig. 4). In Fig. 4a–f we show simultaneous images of $g(\mathbf{r}, E)$ modulations measured at $T = 1.9$ K for six energies near $E_F$ within the energy scale where the resonant feature appears below $T_o$ (Fig. 3b). In Fig. 4g–l we show the six $g(\mathbf{q}, E)$ Fourier transforms of each $g(\mathbf{r}, E)$ image from Fig. 4a–f (Supplementary Fig. 5). Four significant advances are already apparent in these unprocessed data. First, the wavevectors of the hidden-order $g(\mathbf{r}, E)$ modulations are dispersing very rapidly (within the narrow energy range of $DOS(E)$ modifications in Fig. 3b); this is directly indicative of heavy fermions in the hidden-order state. Second, the magnitude of their characteristic **q** vectors, which are diminishing towards a small value as they pass through the Fermi energy from below (see Fig. 4i), suddenly jumps to a large value at a few millielectronvolts above $E_F$ (see Fig. 4k). It therefore appears that the band supporting QPI is widely split in **k**-space at this energy centred a few millielectronvolts above $E_F$. Third, the QPI oscillations are highly anisotropic in **q**-space (Fig. 4g–l). Finally, the most intense modulations rotate by 45° when they pass the energy (compare Fig. 4h and l), indicating a distinct **k**-space electronic structure for the filled and empty gap-edge states. As we show further in Fig. 5, all of these effects are characteristics of the hidden-order state.

## Evolution of k-space structure from Fano lattice to hidden-order state

To determine the **k**-space electronic structure evolution into the hidden-order state, we measure the temperature dependence of QPI data equivalent to those in Fig. 4 from just above $T_o$ down to 1.7 K. In Supplementary Fig. 6 we show the complete temperature dependence of the dispersions of the most intense QPI modulations. The key results are shown in Fig. 5a–d (related $g(\mathbf{q}, E)$ movies are shown in the Supplementary Information) with the relevant **q**-space directions indicated by the blue and red lines on Fig. 4g. With falling temperature below $T_o$, we observe the rapid splitting of a light band (crossing $E_F$ near $(0, \pm 0.3)\pi/a_0$; $(\pm 0.3, 0)\pi/a_0$) into two far heavier bands which become well separated in **k**-space and with quite different anisotropies. The hybridization energy range as estimated from the observed gap at the avoided crossing (see Fig. 1c) is shown by horizontal arrows in Fig. 5 and appears anisotropic in **k**-space by a factor of about two (Fig. 5c, d). This **k**-space structure can also be modelled using equation (2) (Supplementary Fig. 7). Finally, the DOS($E$) changes detected in **r**-space (Fig. 3) occur within the same narrow energy range and, moreover, are consistent with the gaps deduced from thermodynamics and other spectroscopies (Supplementary section VIII).

## Absence of conventional density-wave states

Indications of a conventional density wave would include an energy gap that spans $E_F$, modulations at fixed **Q*** in topographic images, and modulations at fixed **Q*** that are the same for empty and filled gap-edge states in $g(\mathbf{r}, E)$. Searches for all these phenomena, which must occur if the hidden-order state is a conventional density wave with static wavevector **Q*** (ref. 35), were carried out. First, high-precision topographic images of both Si-terminated and U-terminated surfaces are acquired and analysed over the same range of temperatures as in Fig. 3 searching for any additional, bias-independent, modulation wavevectors **Q*** appearing below $T_o$ in the Fourier transform of the topograph. Second, we analyse all the $g(\mathbf{r}, E)$, in search of modulations at fixed $\mathbf{Q}^*$ which are the same for empty and filled gap-edge states. Third, we consider the energy gap structure in $\mathbf{k}$-space revealed by Fig. 5c and d. Because these signatures are not observed at any temperature below $T_o$ for any topographs or $g(\mathbf{r}, E)$ maps (Supplementary Fig. 2), and because the observed $\mathbf{k}$-space alterations do not result in a gap spanning $E_F$ (Fig. 5c, d), we conclude that a conventional density wave does not appear to be present at $T < T_o$ in $URu_2Si_2$.

## Discussion and conclusions

New perspectives on both the general physics of a Kondo lattice (above $T_o$) and the emergence of the hidden order from this state (below $T_o$) are revealed by our SI-STM studies of $URu_2Si_2$. First, Kondo screening of the U-atom magnetism appears to begin below 55 K, a deduction now supported by direct imaging of the expected Fano lattice signature[12–14]. Here the QPI observations in Figs 4 and 5 revealing a light hole-like band with $m^* \approx 8m_e$ cannot account for the $m^* \approx 50m_e$ deduced from specific heat measurements at $T > T_o$; resolution of this issue, which is not relevant to our research objectives here, will require future study. Nevertheless, below $T_o$ we observe this band splitting into two new heavy bands, one of which crosses $E_F$ with $m^* \approx 53$ in the (1, 0) direction and $m^* \approx 17$ in the (1, 1) direction (Fig. 5). Using the geometric mean to estimate the average over $\mathbf{k}$-space, we find $m^* \approx 28m_e$ from our QPI data (Figs 4, 5) while both thermodynamics[15,16] and ARPES[26] report $m^* \approx 25m_e$ for $T \ll T_o$. In addition, we detect the spectroscopic signature of the transition temperature to within 1 K of the bulk value[46] (Fig. 3b), while the partial gap in the tunnelling density of states (Fig. 3b, d) agrees quantitatively with that derived from bulk specific heat studies[15,16,46] (Supplementary Information section VIII). Thus, there is excellent consistency between the electronic structure of the hidden-order state deduced from thermodynamics, photoemission and our new SI-STM approach. Second, if the hidden order below $T_o$ were a conventional density wave, non-dispersive modulations at $\mathbf{Q}^*$ should appear in the gap-edge states both above and below $E_F$ in $g(\mathbf{q}, E)$ and in the topographs, but no such phenomena are detected. Moreover, because the high spectral weight gap-edge states of the hidden-order phase (Figs 3 and 4) are at completely different $\mathbf{k}$-space locations below and above $E_F$ (Figs 4 and 5) and the indirect gap[3] does not cross $E_F$, these effects are also exclusive of a conventional density wave. Third, the actual evolution of electronic structure corresponds to a light dispersing band crossing $E_F$ near $\mathbf{k} = 0.3(\pi/a_0)$ above $T_o$ that undergoes rapid temperature changes below $T_o$. The result is it splits into two far heavier bands that are widely separated in $\mathbf{k}$-space (Figs 4g, 4l, 5c and 5d) within an energy range coincident with changes in the Fano spectrum near $E_F$ (Fig. 3). The dramatic alterations of electronic characteristics[15–26] therefore occur upon the anisotropic heavy-band splitting emerging from the Fano lattice electronic structure. Thus it seems that some sudden modification of the many-body state[12,30–34] periodic with the U atoms (Fig. 2c–e, 3), along with associated alterations to the $\mathbf{r}$-space/$\mathbf{k}$-space hybridization processes (Figs 4 and 5), is the origin of the hidden order of $URu_2Si_2$. Further development of theoretical models rich enough to capture the complex dual $\mathbf{r}$-space/$\mathbf{k}$-space nature of the hidden order is now required.

More generally, the Fano lattice electronic structure images $\varepsilon_0(\mathbf{r})$, $\Gamma(\mathbf{r})$, and $\zeta(\mathbf{r})$ (Fig. 2c–e) allow the first direct visualization of a Kondo-screening many-body state in magnetic lattice (here with $\langle \varepsilon_0 \rangle \approx 3$–4 meV, centred at the U atoms[12]), while quasiparticle interference is used to determine a heavy fermion band structure (both above and below $E_F$). This combined capability to simultaneously visualize the $\mathbf{r}$-space Fano lattice and the $\mathbf{k}$-space heavy fermion structure opens a completely new experimental window onto the physics of multichannel Kondo lattices and heavy fermion physics.

## METHODS SUMMARY

Both pure $URu_2Si_2$ and $U_{0.99}Th_{0.01}Ru_2Si_2$ samples were grown by the Czochralski method, using a continuous gettered tri-arc furnace under Ar gas starting from stoichiometric amounts of the constituent materials. Laue X-ray diffraction measurements were performed for orientation and to check the single-crystal nature. Our 1 mm × 1 mm × 0.5 mm $c$-axis normal samples are inserted into a variable-temperature SI-STM system, mechanically cleaved in cryogenic ultrahigh vacuum, and then inserted into the scanning tunnelling microscope head. Atomically flat and clean $a$–$b$ surfaces consisting of layers of either Si or U atoms (see Figs 2 and 3) are achieved throughout. The cryogenic ultrahigh vacuum conditions enable many months of $g(\mathbf{r}, E)$ measurements with atomic resolution and register in the same FOV without any degradation of surface quality. The SI-STM system was stabilized at each operational temperature between 1.7 K and 29 K for a period of about 24 hours before each set of SI-STM measurements. The $dI/dV(\mathbf{r}, V)$ measurements were acquired with a standard alternating current lock-in amplifier technique, using bias modulations down to 250 μV RMS at the lowest temperatures. The **r**-space measurements on the Si surface required sub-atomic resolution on $20 \times 20$ nm$^2$ FOV while measurements for **k**-space determination used $60 \times 60$ nm$^2$ FOV to ensure high Fourier resolution of the dispersive scattering interference **q**-vectors.

**Acknowledgements** We acknowledge and thank E. Abrahams, M. Aronson, D. Bonn, W. Buyers, A. Chantis, M. Crommie, P. Coleman, D. M. Eigler, M. Graf, A. Greene, K. Haule, C. Hooley, G. Kotliar, D.-H. Lee, A. J. Leggett, B. Maple, F. Steglich, V. Madhavan, A. P. Mackenzie, S. Sachdev, A. Schofield, T. Senthil and D. Pines for discussions and communications. These studies were supported by the US Department of Energy, Office of Basic Energy Sciences, under Award Number DE-2009-BNL-PM015. Research at McMaster University was supported by NSERC and CIFAR. Research at Los Alamos was supported in part by the Center for Integrated Nanotechnology, a US Department of Energy Office of Basic Energy Sciences user facility, under contract DE-AC52-06NA25396 by LDRD funds. P.W. acknowledges support from the Humboldt Foundation, F.M. from the German Academic Exchange Service, and A.R.S. from the US Army Research Office. J.C.D. gratefully acknowledges the hospitality and support of the Physics and Astronomy Department at the University of British Columbia.

Correspondence and requests for materials should be addressed to J.C.D. (jcdavis@ccmr.cornell.edu).

**Figure 1 A Kondo lattice model and its resulting band structure. a**, A schematic representation of the screening of a localized spin-half state (red) by delocalized **k**-space electrons (green) caused by the Kondo effect[1–5]. **b**, A typical Fano tunnelling-conductance spectrum[5] expected near the electronic many-body state depicted in **a**. **c**, A schematic representation of the $T \approx 0$ band structure expected of a Kondo lattice as in equation (2), with the light hole-like band at high temperature depicted by a dashed line. The approximate hybridization energy range is shown by horizontal arrows.

**Figure 2 Imaging the Fano lattice in URu$_2$Si$_2$. a**, A typical topographic image of the Si-terminated surface of URu$_2$Si$_2$. The Si site is marked with a cross and the U site with an X. Data were acquired at −60 mV and 2 GΩ junction resistance. **b**, A typical spatially averaged Fano-like $<g(E)>$ spectrum detected on all Si-terminated surfaces of URu$_2$Si$_2$ at $T < 20$ K. The inset shows the layered structure of the crystal with the U-terminated surface; the Si-terminated surface is two atomic layers below with each Si at the middle site between four U atoms. **c**, Image of the many-body state energy $\varepsilon_0(\mathbf{r})$ extracted from fitting the spatially resolved Fano spectrum according to equation (1); the FOV is indicated by the yellow box in **a**. U atoms are designated by an X and the maximum in $\varepsilon_0$ always occurs at these sites. **d**, Image of the hybridization width $\Gamma(\mathbf{r})$ extracted from fitting the spatially resolved Fano spectrum according to equation (1); the FOV is the same as in **c**. The minimum in $\Gamma$ occurs at the U sites. **e**, Image of the ratio of electron tunnelling probability $\zeta(\mathbf{r})$ extracted from fitting the spatially resolved Fano spectrum according to equation (1); the FOV is the same as in **d** and **e**. The maximum in $\zeta$ occurs at the U sites.

**Figure 3 Evolution of DOS(E) upon entering the hidden-order phase. a**, Topographic image of U-terminated surface with the temperature dependence of its spatially averaged spectra $<g(E)>$ in the inset. Each of these spectra is shifted vertically by 5 nS for clarity. Blue data are within 1 K of $T_o$ for 1% Th-doped samples. The image was taken at −10 mV and 2.5 MΩ junction resistance. **b**, Temperature dependence of DOS($E$) modifications due to the appearance of the hidden order at the U-terminated surfaces. Each spectrum is derived by subtracting the spectrum for $T > T_o$ (and shifted vertically for clarity). The DOS($E$) changes are limited to approximately ±5 meV. **c**, Topographic image of Si-terminated surface with the temperature dependence of its spatially averaged spectra $<g(E)>$ in the inset. Each of these spectra is shifted

vertically by 5 nS for clarity. The image was taken at 150 mV and 3 GΩ junction resistance. **d**, Temperature dependence of DOS($E$) modifications due to the appearance of the hidden order at the Si-terminated surfaces. Each spectrum is derived by subtracting the fit to a Fano spectrum (equation (1)), which excludes data points in the range −7.75 mV to 6.75 mV. The DOS($E$) changes are again limited to approximately ±5 meV.

**Figure 4 Energy dependence of heavy *f*-electron quasiparticle interference. a–f**, Atomically resolved $g(\mathbf{r}, E)$ for six energies measured at the U-terminated surface. Extremely rapid changes in the interference patterns occur within an energy range of only a few millielectronvolts. Data were acquired at –6 mV and 25 MΩ setpoint junction resistance. **g–l**, Fourier transforms $g(\mathbf{q}, E)$ of the $g(\mathbf{r}, E)$ in **a–f**. The associated $g(\mathbf{q}, E)$ movie is shown in the Supplementary Information. The length of half-reciprocal unit-cell vectors are shown as dots at the edge of each image. Starting at energies below $E_\mathrm{F}$ (**g**), the predominant QPI wavevectors diminish very rapidly until **i**; upon crossing a few millielectronvolts above $E_\mathrm{F}$, they jump to a significantly larger value and rotate through 45°. Then they again diminish in radius with increasing energy in **j**, **k** and **l**. This evolution is not consistent with a fixed $\mathbf{Q}^*$ conventional density wave state but is consistent with an avoided crossing between a light band and a very heavy band.

**Figure 5 Emergence of the two new heavy bands below the hidden-order transition. a**, Dispersion of the primary QPI wavevector for $T > T_\mathrm{o}$ along the (0, 1) direction (see Fig. 4g). A single light hole-like band crosses $E_\mathrm{F}$. **b**, Dispersion of the primary QPI wavevector for $T > T_\mathrm{o}$ along the (1, 1) direction (see Fig. 4g). A single light hole-like band crosses $E_\mathrm{F}$. **c**, Dispersion of the primary QPI wavevector for $T \approx 5.9$ K along the (0, 1) direction (see Fig. 4g). Two heavy bands have evolved from the light band and become well segregated in **k**-space within the hybridization gap. **d**, Dispersion of the primary QPI wavevector for $T \approx 5.9$ K along the (1, 1) direction (see Fig. 4g). Two heavy bands have evolved from the light band and are again segregated in **k**-space within the hybridization gap.

Figure 1

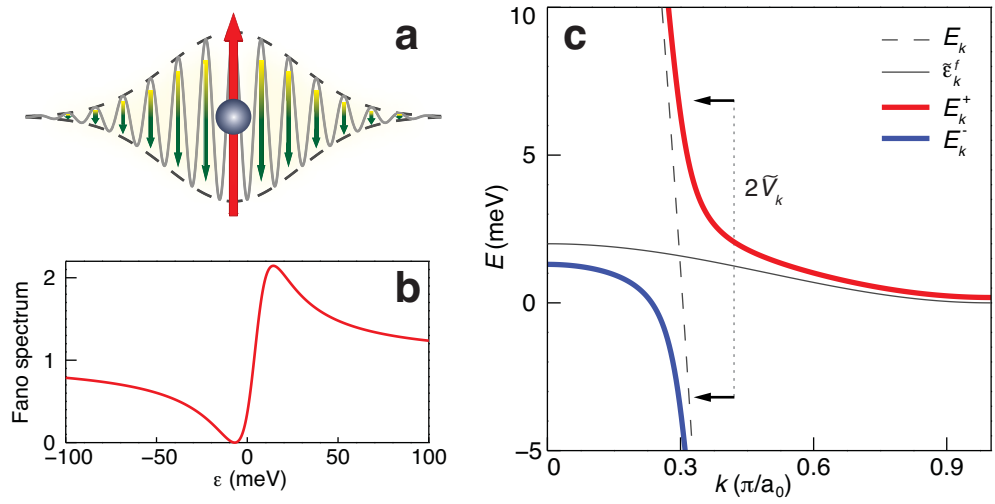

Figure 2

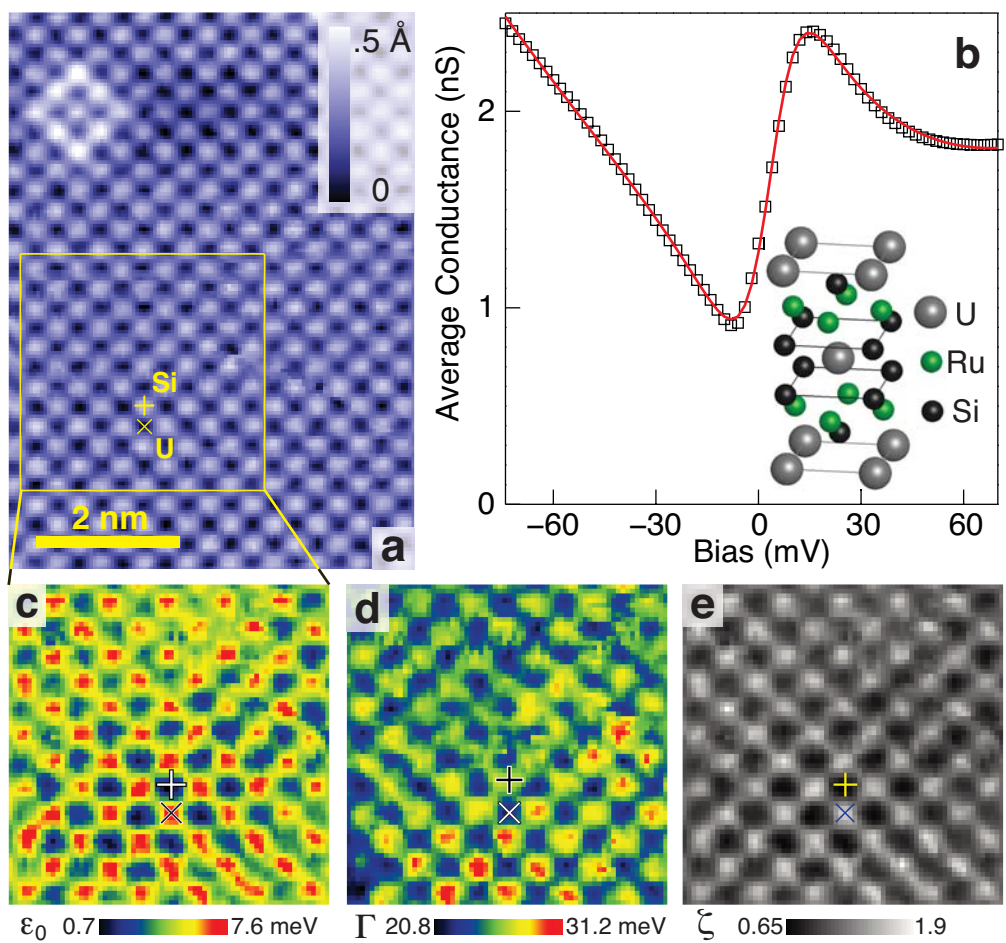

Figure 3

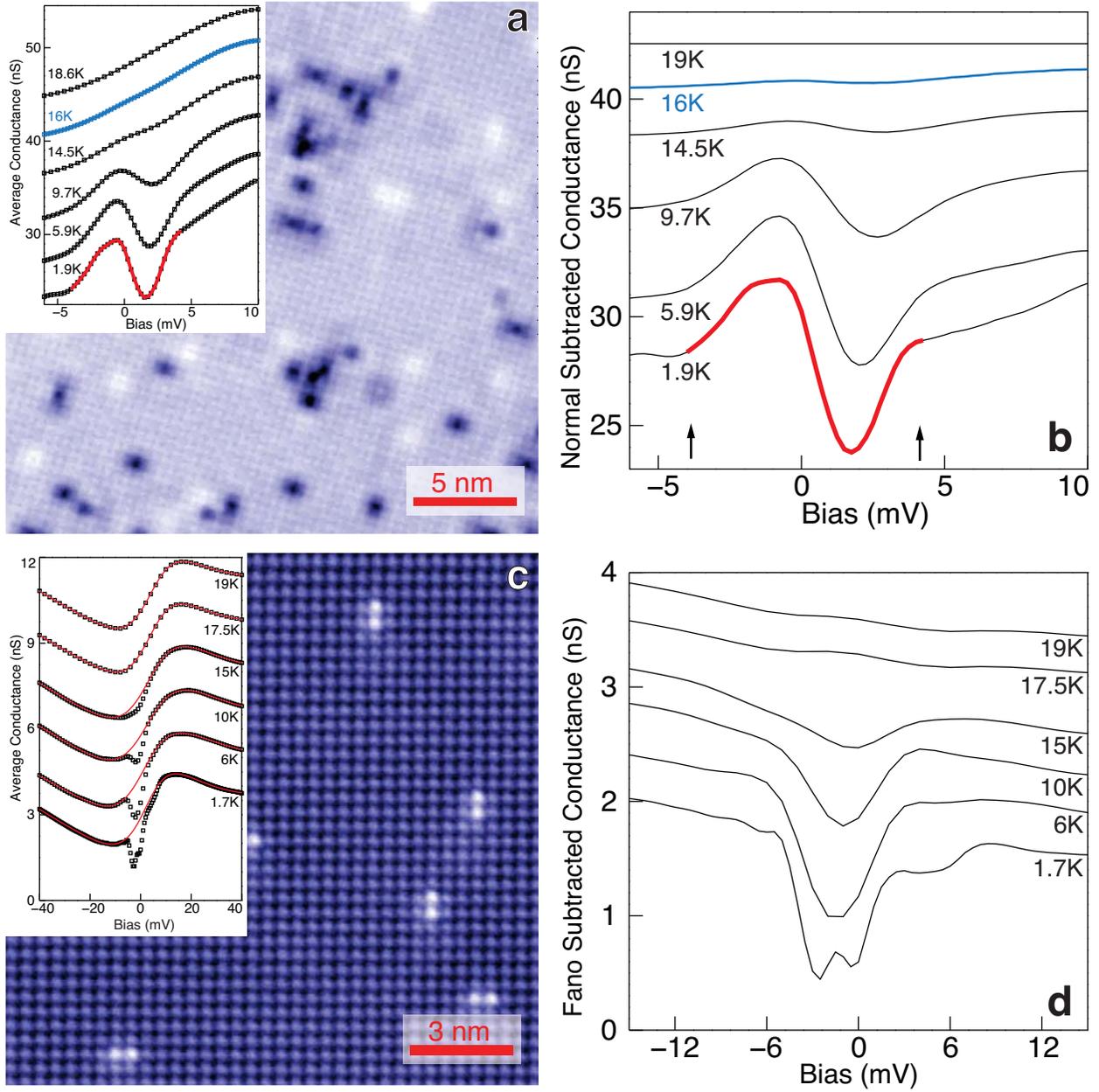

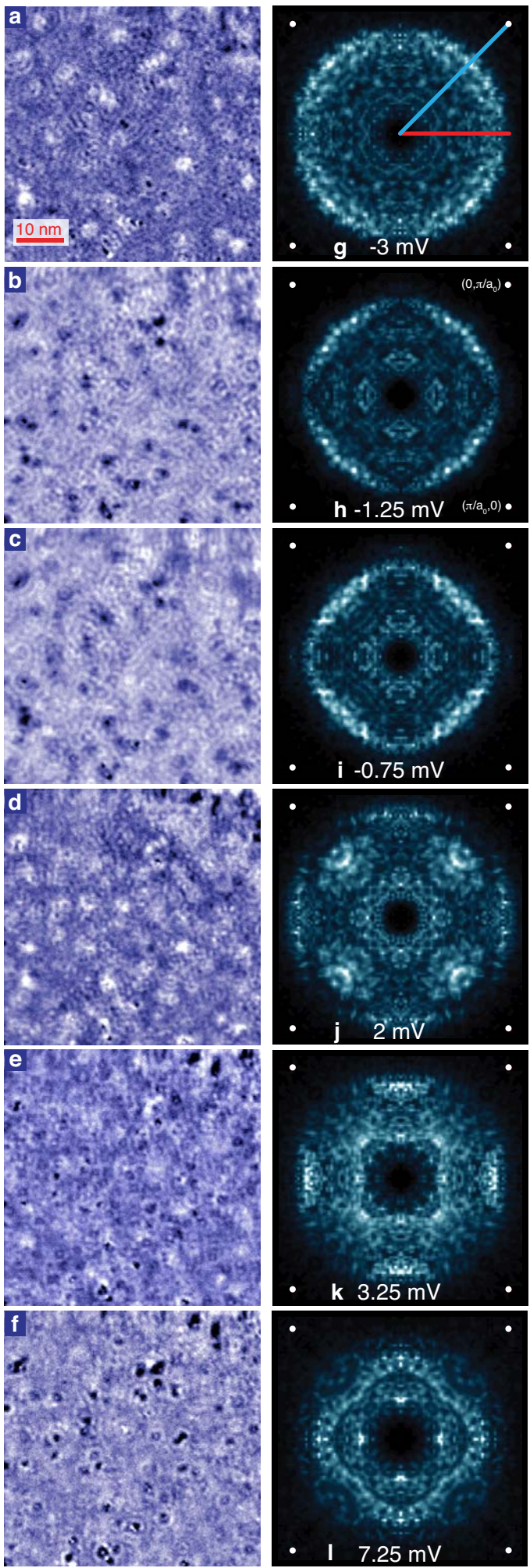

Figure 4

Figure 5.

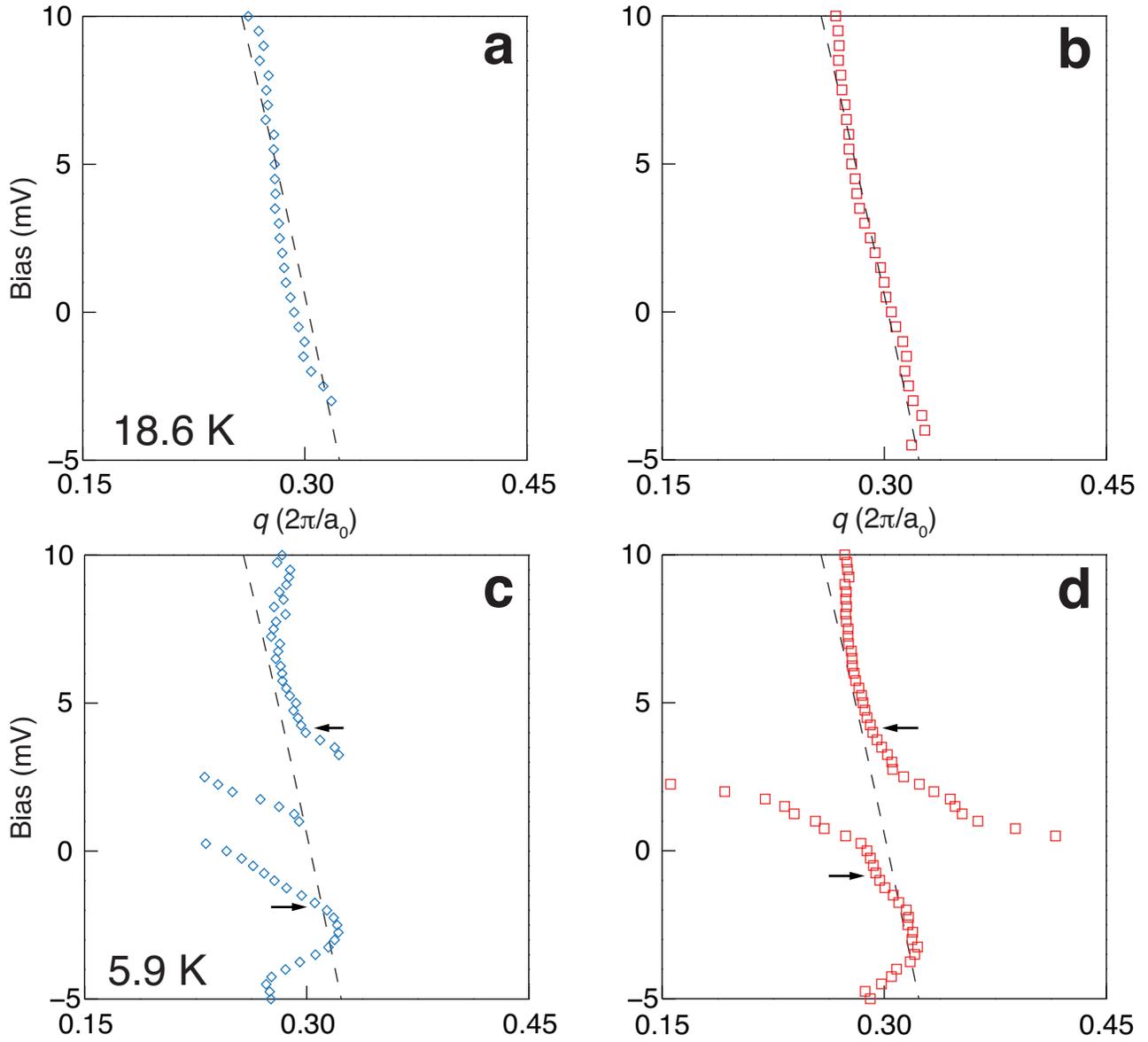

Supporting Online Material for

**Emergence of the Hidden Order State from the Fano Lattice Electronic Structure of the Heavy-fermion Material $URu_2Si_2$**

A.R. Schmidt, M.H. Hamidian, P. Wahl, F. Meier, A.V. Balatsky, T. J. Williams, G.M. Luke and J.C. Davis*

## (I) Sample Preparation

Both the pure $URu_2Si_2$ and the $U_{0.99}Th_{0.01}Ru_2Si_2$ samples were grown at the McMaster University Brockhouse Institute for Materials Research by the Czochralski method, using a continuous gettered tri-arc furnace under Ar gas starting from stoichiometric amounts of the constituent materials. No additional heat treatment was done on these as-grown crystals. Laue x-ray diffraction measurements were performed for orientation and to check the single-crystal nature. The samples were cut by spark erosion along the desired direction.

## (II) SI-STM Procedures

In SI-STM, measurement of the STM tip-sample differential tunneling conductance $dI/dV(\mathbf{r},V) \equiv g(\mathbf{r},E=eV)$ at locations $\mathbf{r}$ and sample-bias voltage $V$, yields an image proportional to the local density of electronic states $LDOS(\mathbf{r},E)$. Moreover, $\mathbf{k}$-space electronic structure elements can be determined simultaneously by using Fourier transform scanning tunneling spectroscopy (FT-STS). This is because the spatial modulations in $g(\mathbf{r},E)$ due to interference of quasiparticles scattered by atomic-scale impurities is detectable in $g(\mathbf{q},E)$, the Fourier transform of $g(\mathbf{r},E)$. We use these techniques to examine simultaneously the $\mathbf{r}$-space $\mathbf{k}$-space electronic structure evolutions in $URu_2Si_2$ and $U_{0.99}Th_{0.01}Ru_2Si_2$. The 1mm x 1mm x 0.5mm c-axis normal samples are inserted into a variable temperature SI-STM system, mechanically cleaved in cryogenic ultra-high vacuum, and then inserted into the STM head. Atomically flat and clean a-b surfaces consisting of layers of either Si or U atoms (see inset Fig. 2) are achieved throughout. The cryogenic UHV conditions enable many months of measurements registered to the same atoms on the same surface without any degradation of surface quality. The SI-STM system was stabilized at each operational temperature between 1.7K and 29K for a period of ~24 hours prior to all tunneling measurements. The differential conductance measurements were acquired with a standard AC lock-in amplifier technique using bias modulations down to 0.25mV RMS at lowest temperatures. Real-space measurements on the Si surface required sub-atomic resolution on 20nm² field of view (FOV). The measurements for *k*-space studies use 60nm² FOV to ensure high Fourier resolution of the dispersing scattering **q**-vectors.

## (III) Fano Fit Procedures

Above the hidden order transition temperature (17.5K), the measured LDOS spectrum at every point of the Si surface can be fitted with high quality to a Fano function plus quadratic background:

$$g(E=eV) \propto \frac{(q+E')^2}{E'^2+1} + aV^2 + bV + c; \quad E' = \frac{(V-\varepsilon_0)}{\Gamma/2}$$

To account for thermal broadening, in the fits this function was convolved with the derivative of the Fermi-Dirac distribution at every temperature. By varying the bias voltage interval over which the data is fit from [-40, 40mV] to [-75, 75mV], the Fano parameter lattice presented in Figure 2 does not change qualitatively. Quantitatively, the mean values of $\varepsilon_0$ and $\Gamma$ shift by ~1.5 meV and ~3 meV respectively, and $q$ shifts by ~0.2. The change of the standard deviation of these parameters is less than 5%, so that the systematic effect of varying the bias interval over which the fit is performed is a primarily a rigid shift of Fano parameters, maintaining the shapes of their distribution. Examples of fits to a single spectrum, above 17.5K, at a Si site and at a U site are shown in Figure S1.

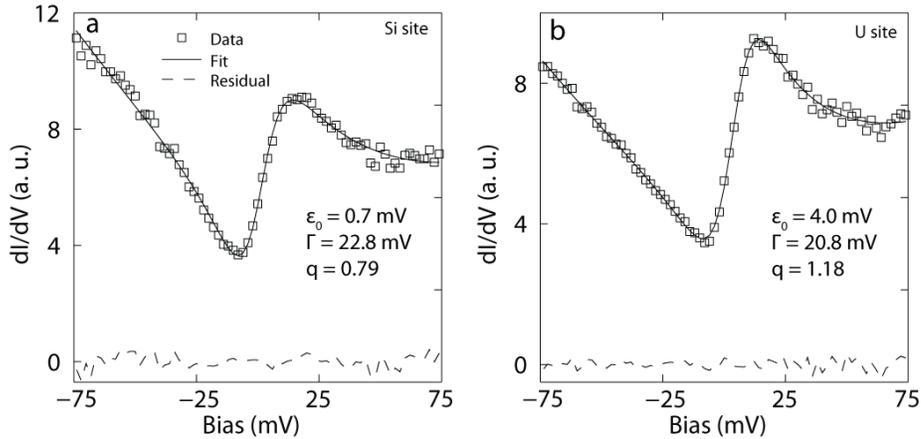

**Fig S1:** Resulting fits of Fano function to individual measured spectra on a (a) Si site and a (b) U site for T = 19K.

Below 17.5K, a dip is observed to open in the LDOS spectrum in the vicinity of $\varepsilon_0$. This dip does not appear to be due to a drastic change in the Fano parameters with temperature but instead appears to be a non-Fano gap opening over a larger Fano lineshape that is not changing significantly with temperature. To test this hypothesis, we fit the thermally broadened Fano function above at every temperature excluding the bias interval [-7.75, 6.75]. This produces the red lines over the data inset in figure 2c. The excellent agreement of these fits with the excluded data at 17.5K and 19K confirms the internal consistency of this exclusion procedure.

## (IV) Absence of Conventional Density Wave

Conventional density wave ordering has long been a contending postulate for the hidden order phase [26,31]. Our data directly demonstrates its absence or at the very least restricts the upper bounds of the associated charge moment to extremely low values. STM topographs, which are images of the spatially resolved energy integrated electronic local density of states (LDOS), are sensitive to charge modulations beyond those simply attributed to the lattice corrugation. Consequently, the STM is enabled to directly detect charge density waves[38]. Power spectral density (PSD) profiles of temperature dependent topographs obtained on both Si and U terminated surfaces are presented in figures S2 (a)-(f). Immediately obvious are the reciprocal lattice peaks present in all the panels. Were a charge density wave present, the images would boast an additional set of wavevector peaks. Furthermore, a conventional density wave modulation of the atomic lattice would generate a beat pattern of the reciprocal lattice peaks in the PSD, akin to the effect of the supermodulation in BSCCO [39]. By spanning temperatures across $T_0$, the data clearly demonstrates the absence of any additional features appearing below the hidden order transition.

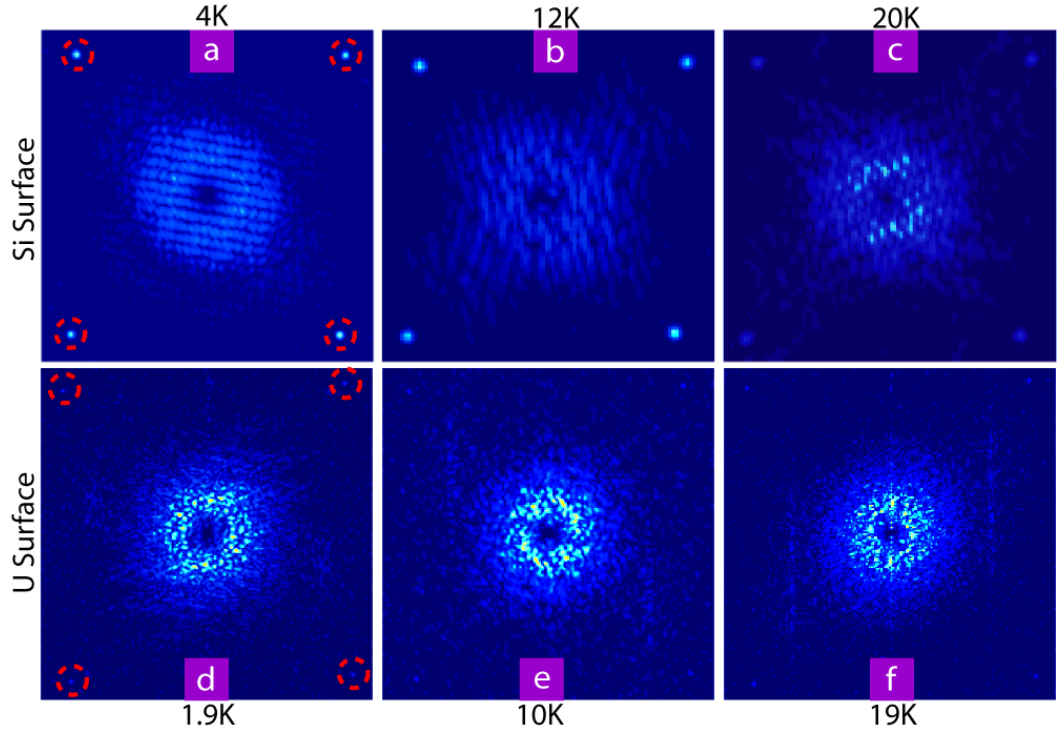

**Fig. S2:** FT of the U-layer (a)-(c) and Si-layer (d)-(f) topographs across temperatures above and below the hidden order transition. The dotted red circles are a guide to find the regions of the reciprocal lattice vectors in each of the panels.

## (V) Quasiparticle Interference Imaging on the Si Surface

QPI extraction techniques, as described in part VII, applied to conductance maps on the U-layer of $URu_2Si_2$ revealed Kondo-like dispersions in the electronic band structure below the hidden order transition. Identifying the bands can be non-trivial or practically impossible without additional information or assumptions about the underlying band structure. In the case of the U-layer, the assumption of isotropic single-band scattering, implemented through the relation $q = 2k$, within an energy range of ±10mV about the chemical potential, pulls out the Kondo-like hybridized momentum space band from the ***q***-space wavevectors. Hence, the success of this disentanglement rests upon being able to clearly identify dispersing QPI wavevectors and to also devise the correct scattering vectors which relates the ***k***- space structure with the ***q***-space structure.

Data collected on the Si-layer has so far proved unyielding to QPI analysis. The complex ***q***-space structure, as seen in Fig. S3, makes it very difficult to find the dispersing wavevectors without which one cannot devise the transformation between momentum and ***q***-space. Even along the high symmetry direction of (1,0) and (1,1), linecuts show a complicated lineshape with possibly multiple dispersive wavevectors. In comparison, Fig. 3 (g)-(l) of the main text demonstrates a beautifully simple set of dispersing wavevectors.

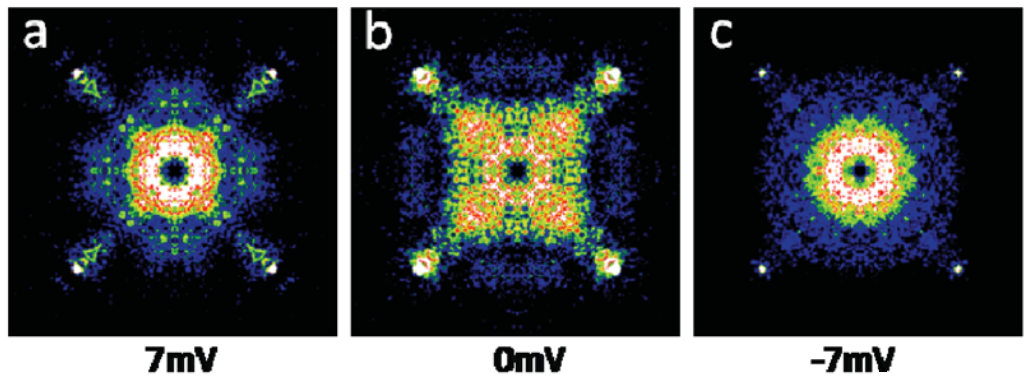

**Fig. S3:** *q*-space representation of conductance maps on the Si-layer of $URu_2Si_2$ for three different measurement biases. The complex structure makes identifying and tracking QPI wavevectors difficult.

### (VI) Simultaneous Topography of U Layer Data in Fig. 3

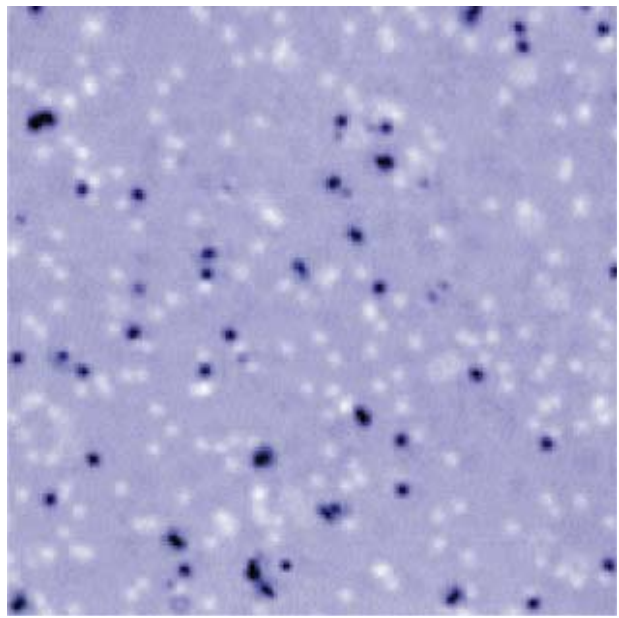

**Fig. S4:** Topograph of the 45nmx45nm FOV on which the conductance map represented in Fig. 3 of the main text was taken. Dark spots correspond to surface Th atoms. White spots correspond to subsurface impurities.

## (VII) Extraction of QPI Modulation Wavevectors via Power Spectral Analysis

Real space modulations observed in the conductance maps resulting from quasiparticle interference, are made manifestly clear by utility of Fourier transform techniques. To enhance the signal to noise ratio (S/N) of the modulation vectors in the conjugate representation ($q$-space), additional steps are employed. Starting from the 2-D PSD of the conductance data, linear transformations are applied to restore the lattice symmetry that may have been distorted due to piezo relaxation effects of the STM actuators. Each energy layer is then symmetrized along the high crystalline symmetry directions (1,0) and (1,1) yielding enhanced visibility of the QPI wavevectors. Fig. S5 makes evident the significant improvement in S/N but with no change in the fundamental content of the data, by applying this procedure. The dominant QPI modulation wavevectors associated with the onset of hidden order are extracted from the $q$-space representation of the data by performing linecuts, with transverse multi-pixel averaging, along the (1,0) and (1,1) directions. By applying a Lorentzian fit, the magnitude of the QPI vector, $q = |q|$, can be determined with great precision and its energy dispersion traced as the procedure is applied to each energy layer of the conductance map.

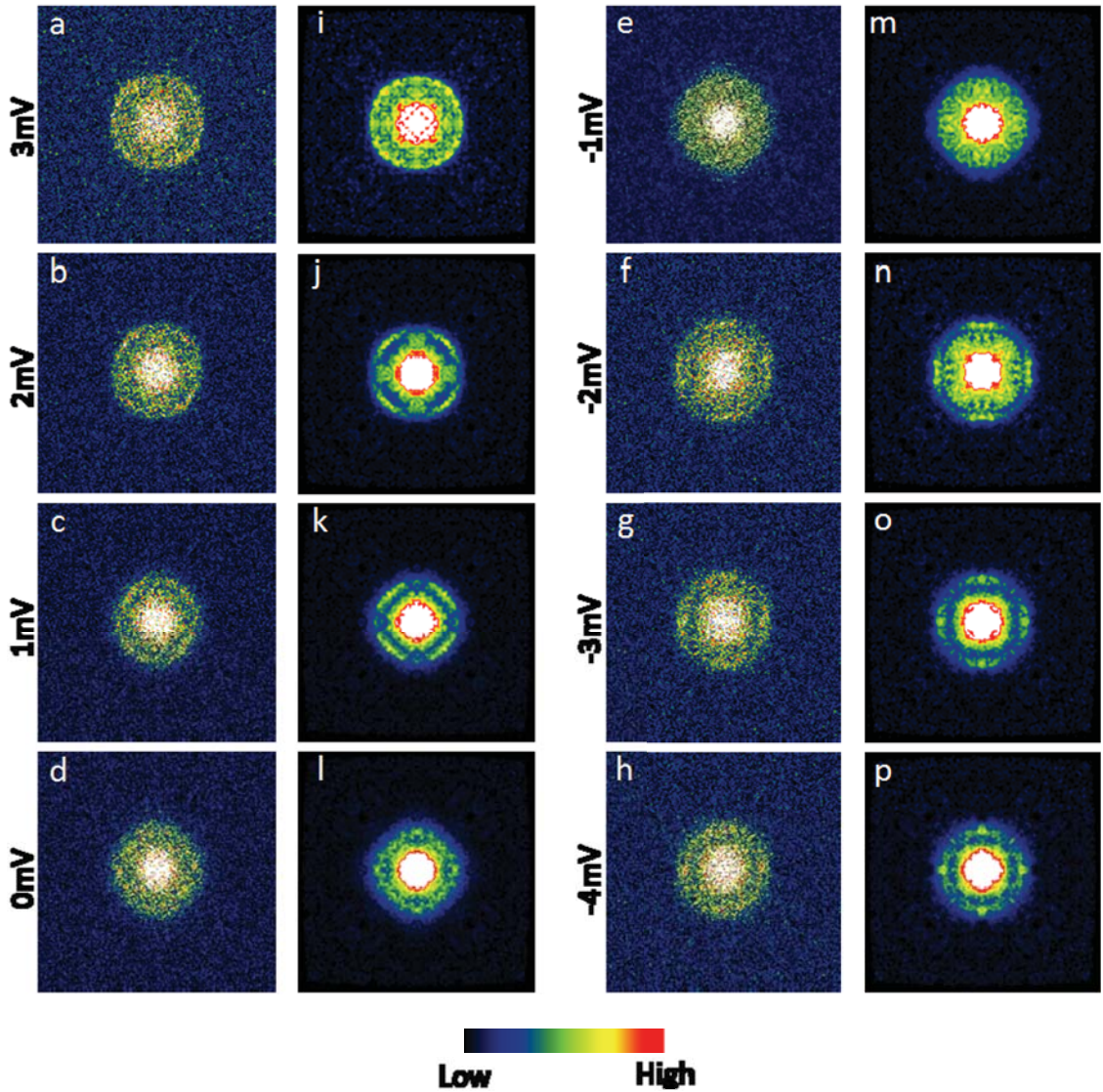

**Fig. S5:** Comparison between raw *q*-space data (a)-(h), and *q*-space data treated with the procedure described in section B,(i)-(p).

### (VIII) Temperature Dependence of QPI Modulation Vectors

Figure 4 of the main text plots the wavevector magnitudes of the highest amplitude modulations for two temperatures, one just above the hidden order transition and the other well below. The anisotropy below the hidden order transition is demonstrated as the direction of the most intense modulations rotates by 45 degrees passing through an energy just above $E_F$. Figure S6 plots the detailed temperature evolution of the complete set of observed QPI wavevector magnitudes along the high symmetry directions (1,0) and (1,1). The rapid splitting of the bands participating in the quasiparticle scattering processes is observed as the temperature is lowered. Without the spectral intensity information, bands along (1,0) and (1,1) exhibit similar evolutions down to ~6K where additional anisotropy, beyond the already observed spectral weight distribution, begins to form.

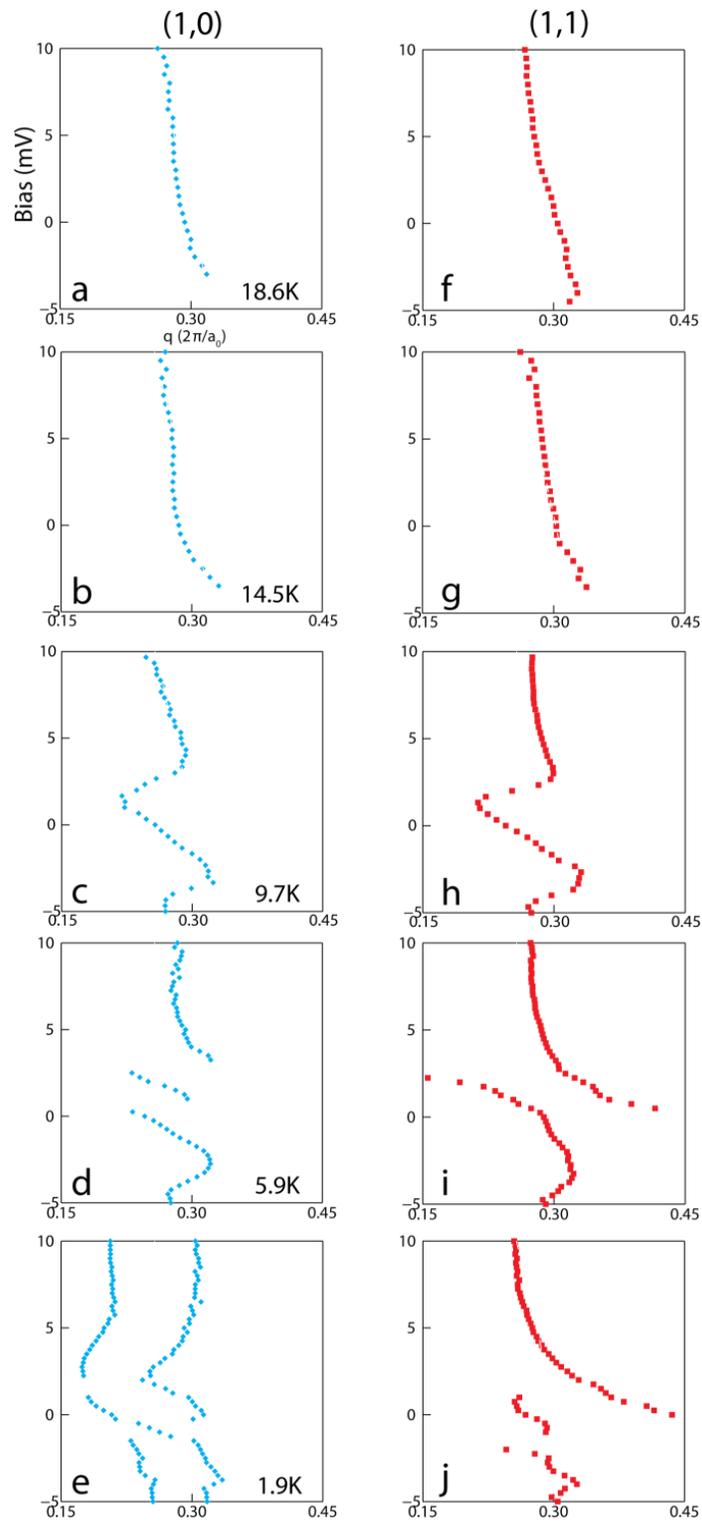

**Fig. S6:** Complete temperature dependence of the full set of observed QPI wavevectors for the two high symmetry directions (1,0) and (1,1) at a series of temperatures as shown inset to panels a-e.

# (IX) Comparing STM and Thermodynamic Deduced Gap Features

Figures 2 (b) and (d) of the main text render the emergence of gap like features in the density of states revealed through measurements on the U surface of $Th_{0.01}U_{0.99}Ru_2Si_2$ and the Si surface of $URu_2Si_2$, respectively. Specific heat studies of these compounds [22,37] yield inferred gap widths are in relatively good agreement with our values. Figure S7(a) shows that the DOS of states gap observed on the Si surface of the undoped sample is approximately 10mV while the specific heat measurements of reference 21 give an estimate of 11mV. Similarly, the reduced gap size of 4.2mV in the 2% Th-diluted compound reported by Ref. 36 is in accord with the width of DOS gap like feature depicted in the figure S7(b).

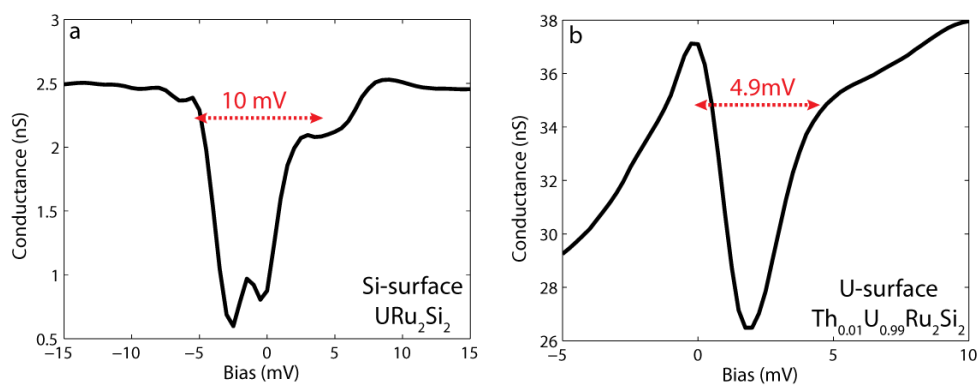

**Fig. S7:** (a) Extracted gap feature on the Si surface of $URu_2Si_2$. (b) Characteristic gap like feature on the U surface of $Th_{0.01}U_{0.99}Ru_2Si_2$.

Thus, the specific heat anomaly which is directly associated with the hidden order transition implies a gapping of the Fermi surface which is in strong quantitative agreement with the DOS gap features observed by STM. The link provides direct evidence that the DOS gap structures and the directly associated band splitting, observed through QPI, are the microscopic electronic signatures of the second order phase transition.